\documentclass[10pt,aps,prl,superscriptaddress,twocolumn,preprintnumbers,nofootinbib]{revtex4-2}
\usepackage{amsmath,amssymb,mathtools}
\usepackage{booktabs}
\usepackage{graphicx}
\usepackage{lineno}
\usepackage{hyperref}
\usepackage{color}
\usepackage{comment}

\hypersetup{
  colorlinks=true,
  linkcolor=blue,
  citecolor=blue,
  urlcolor=blue,
}

\newcommand{\ycut}{y_{\mathrm{cut}}}
\newcommand{\alphas}{\alpha_{\mathrm{s}}}

\begin{document}
	

\preprint{CERN-TH-2026-017, IPPP/26/12, ZU-TH 04/26, LAPTH-007/26}

\title{The Four-Jet Rate in Electron-Positron Annihilation at Order $\alpha_s^4$}

\author{Xuan Chen}
\affiliation{%
School of Physics, Shandong University, Jinan, Shandong 250100, China}%

\author{Dmitry Chicherin}
\affiliation{%
LAPTh-CNRS-USMB, 9 chemin de Bellevue, 74940, Annecy, France}%

\author{Elliot Fox}
\affiliation{%
Institute for Particle Physics Phenomenology, Department of Physics, University of Durham, Durham, DH1 3LE, UK}%

\author{Nigel Glover}
\affiliation{%
Institute for Particle Physics Phenomenology, Department of Physics, University of Durham, Durham, DH1 3LE, UK}%

\affiliation{%
Theoretical Physics Department, CERN, 1211 Geneva 23, Switzerland}%

\author{\mbox{Matteo Marcoli}}
\affiliation{%
Institute for Particle Physics Phenomenology, Department of Physics, University of Durham, Durham, DH1 3LE, UK}%

\author{Vasily Sotnikov}
\affiliation{%
Physik-Institut, Universit\"at Z\"urich, Winterthurerstrasse 190, 8057 Z\"urich, Switzerland}%

\author{Huiting Sun}
\affiliation{%
School of Physics, Shandong University, Jinan, Shandong 250100, China}%

\author{Hantian Zhang}
\affiliation{%
Theoretical Physics Department, CERN, 1211 Geneva 23, Switzerland}%

\author{Simone Zoia}
\affiliation{%
Physik-Institut, Universit\"at Z\"urich, Winterthurerstrasse 190, 8057 Z\"urich, Switzerland}%
	
\begin{abstract}
 We compute for the first time the production rate for four jets in electron-positron annihilation at next-to-next-to-leading order. Our calculation exhibits the highest final-state jet multiplicity considered at this perturbative accuracy to date.
  The cancellation of infrared singularities is achieved in the antenna subtraction scheme, relying particularly on generalized antenna functions.
  The evaluation of the two-loop virtual corrections is enabled by the construction of a new basis of transcendental special functions tailored to four-particle decay kinematics. 
  Our results are compared with LEP data, finding improved agreement with respect to the next-to-leading order calculation. In the region where perturbative predictions are most reliable, we observe a significant reduction of theory uncertainties, which now fall below the experimental ones. 
\end{abstract}
	
\maketitle

\section{Introduction}

The study of hadronic final states in high-energy particle collisions—particularly jet observables—has been instrumental in establishing Quantum Chromodynamics (QCD) as the theory of the strong interaction. This is especially true in electron–positron annihilation, where the absence of initial-state QCD radiation and underlying-event effects provides a uniquely clean environment for precision studies of jets and event shapes~\cite{Stagnitto:2025air}. Landmark achievements include the discovery of the gluon in planar three-jet events~\cite{TASSO:1979zyf}, the measurement of its spin~\cite{L3:1991jmb}, and stringent tests of the non-Abelian gauge structure of QCD through three- and four-jet production~\cite{L3:1990jlf,ALEPH:1992fwh,ALEPH:1997mcm,Kluth:2000km,OPAL:2001klt,ALEPH:2002kjp,DELPHI:2004als,Kluth:2006bw}. These developments were accompanied by increasingly precise determinations of the strong coupling constant, $\alpha_s$ (see Ref.~\cite{Huston:2023ofk} and references therein).

Precision QCD measurements constituted a central component of the physics programme at the Large Electron–Positron Collider (LEP). Since then, substantial advances have been made in higher-order perturbative techniques, significantly enhancing the accuracy of theoretical predictions and enabling increasingly stringent comparisons with experimental data. Fresh opportunities for QCD studies are now emerging from renewed analyses of LEP data employing modern analysis strategies and theoretical inputs (see, \textit{e.g.}, Ref.~\cite{Electron-PositronAlliance:2019cpi}). Looking ahead, future lepton colliders such as the Future Circular Collider (FCC-ee)~\cite{FCC:2018byv,FCC:2018evy,FCC:2025lpp} and the Circular Electron–Positron Collider (CEPC)~\cite{An:2018dwb,CEPCPhysicsStudyGroup:2022uwl,Ai:2024nmn} are expected to deliver unprecedented experimental precision. Realizing their full physics potential will require commensurate progress in theoretical predictions, in particular in high-order perturbative QCD and related precision frameworks.

While the inclusive cross section for hadron production at electron-positron colliders has been computed to high order~\cite{Gorishnii:1990vf,Surguladze:1990tg,Chetyrkin:1996ela,Baikov:2008jh,Baikov:2010je,Baikov:2012er} in perturbative QCD, less precise predictions are typically available for more differential observables, like event shapes and jet rates. The inclusive $n$-jet rate is defined as the fraction of hadronic events in which $n$ distinct jets are reconstructed according to a given jet clustering algorithm. In the following, we consider the Durham (or $k_T$) jet algorithm~\cite{Catani:1991hj,Brown:1990nm,Brown:1991hx,Stirling:1991ds,Bethke:1991wk}, which has been extensively used for jet studies at lepton colliders. The algorithm proceeds as follows: for each pair $(i,j)$ of final-state particles, the distance measure
\begin{equation}
    y_{ij}=\dfrac{2\min(E_i^2,E_j^2)(1-\cos\theta_{ij})}{Q^2}
\end{equation}
is evaluated, where $Q=\sum_k E_k$ represents the sum of all final-state particle energies $E_k$, and $\theta_{ij}$ is the angle within each pair. The particles with the minimum distance $y_{ij}$ are then clustered together by summing their four momenta. The previous steps are repeated until all pair-wise distances are larger than a pre-defined jet resolution parameter $\ycut$, and the event is classified as an $n$-jet event, where $n$ is the number of remaining final-state clusters. Clearly, the smaller $\ycut$, the more distinct jets are resolved in an event, and vice versa.

Jet rates as a function of $\ycut$ can be computed by normalizing the cross section for the production of $n$-jets in electron-positron annihilation for a given value of $\ycut$, denoted with $\sigma_n(\ycut)$, by the fully-inclusive cross section for hadron production $\sigma_{\text{inc.}}$. The $n$-jet rate then reads
\begin{equation}
\label{eq:jet_rate}
R_n(\ycut) = \frac{\sigma_n(\ycut)}{\sigma_{\text{inc.}}}\,.
\end{equation}
The leading order (LO) contribution to $R_n(\ycut)$ starts at order $\alphas^{n-2}$. Next-to-leading order (NLO) corrections in QCD have been computed for up to seven jets~\cite{Ellis:1980wv,Kunszt:1980vt,Vermaseren:1980qz,Fabricius:1981sx,Campbell:1998nn,Dixon:1997th,Nagy:1997yn,Frederix:2010ne,Becker:2011vg}. Next-to-next-to-leading order (NNLO) predictions for $R_3(\ycut)$ (order $\alphas^3$) were first presented in Refs.~\cite{Gehrmann-DeRidder:2008qsl,Weinzierl:2008iv}. In Ref.~\cite{Gehrmann-DeRidder:2008qsl}, the implementation of the $e^+e^-\to jjj$ process at NNLO in \texttt{EERAD3}~\cite{Gehrmann-DeRidder:2014hxk,Aveleira:2025svg} is combined with the knowledge of the inclusive cross section up to order $\alphas^3$~\cite{Gorishnii:1990vf,Surguladze:1990tg,Chetyrkin:1996ela} to also reach next-to-next-to-next-to-leading order (N$^3$LO) accuracy on $R_2(\ycut)$. In Ref.~\cite{Chen:2025kez}, a direct N$^3$LO calculation of $R_2(\ycut)$, which does not rely on the previous knowledge of the inclusive cross section at order $\alpha_s^3$, was presented. We note that quarks are considered massless in Refs.~\cite{Gehrmann-DeRidder:2008qsl,Weinzierl:2008iv,Chen:2025kez}. Mass corrections to $R_3(\ycut)$ were first computed in Refs.~\cite{Brandenburg:1997pu,Nason:1997nw,Rodrigo:1997gy} at NLO. Beyond fixed-order, all-order resummation effects in jet rates have been computed up to next-to-next-to-leading logarithmic (NNLL) accuracy for $R_2(\ycut)$~\cite{Banfi:2016zlc} and $R_3(\ycut)$~\cite{vanBeekveld:2024wws}, and at next-to-leading logarithmic (NLL) accuracy for higher-multiplicity rates~\cite{Nagy:1998kw,Baberuxki:2019ifp}.

In this Letter, we present the first calculation of $R_4(\ycut)$ at order $\alphas^4$, obtained from an NNLO computation of the four-jet production process $e^+e^- \to jjjj$.
This calculation lies at the current frontier of perturbative QCD.
The explicit cancellation of infrared divergences and the phase-space integration are carried out for the first time at NNLO for such a high final-state jet multiplicity.
Two-loop corrections for five-particle scattering with one leg off shell have only recently begun to emerge~\cite{Abreu:2021asb,DeLaurentis:2025dxw,Badger:2024mir,Badger:2021nhg,Badger:2022ncb,Badger:2021ega,Abreu:2021smk,Abreu:2020jxa,Badger:2024sqv,Abreu:2023rco,Canko:2020ylt,Kardos:2022tpo,Papadopoulos:2015jft,Chicherin:2021dyp}.
In this work, we construct a new basis of transcendental special functions that enables analytic results for such two-loop corrections in the four-particle decay region.
This construction allows us to analytically continue the two-loop amplitudes for hadronic vector-boson production in association with two jets, currently known only in the leading-color approximation (LCA)~\cite{Abreu:2021asb,DeLaurentis:2025dxw}, into the decay region relevant for $e^+e^- \to jjjj$.

Our results represent a crucial step to improve the accuracy of theoretical predictions for electron-positron colliders. 
In particular, we address one of the many challenges~\cite{deBlas:2025gyz} that theoretical calculations must meet in preparation for the next generation of electron-positron colliders.

\section{Calculation}

We perform the calculation within the \texttt{NNLOJET} Monte Carlo framework~\cite{NNLOJET:2025rno}, suitably extended to include four-jet production at electron-positron colliders up to NNLO.

The jet rates that we consider in this work are normalized to the inclusive hadron production cross section, and are integrated inclusively over all angular directions. For such observables the electroweak coupling, the loop corrections to the electroweak boson propagator, and the effects of $\gamma^*/Z$ mixing cancel in the ratio, provided that quarks are treated as massless and that singlet contributions are ignored~\cite{Chetyrkin:1996ela}. We consider five massless quark flavors, $N_f=5$, neglect top-quark loop effects, which are expected to be small~\cite{Dixon:1997th,Campbell:2016tcu}, and we do not  systematically include singlet contributions, which were observed to be numerically negligible at NLO~\cite{Kniehl:1989bb,Dixon:1997th, Hagiwara:1990dx, Signer:1996bf,Frederix:2010ne}, as we also verified explicitly.
In particular, the contributions from anomalous diagrams with axial-vector coupling of the $Z$-boson to closed quark loops cancel between massless isospin doublets,
and we exclude the remaining negligible contribution from bottom-top-quark mass splitting.
Under these assumptions, the results we present are therefore valid for both $\gamma^*$ and $Z$ exchange. In particular, our predictions for $R_4(\ycut)$ depend on the center-of-mass energy $\sqrt{s}$ only via the running of the strong coupling. 

An additional approximation is applied only to the two-loop infrared-finite remainder due to its complexity: we retain only the leading terms in the $N_c \to \infty$ limit, with $N_f/N_c$ constant in each partonic channel (LCA). This approximation has been found to be generally reliable for the double-virtual contributions to five-particle processes provided these contributions are subdominant~\cite{Badger:2023mgf,Abreu:2023bdp,Buccioni:2025bkl,Czakon:2025wgs}. Finally, in the two-loop finite remainder, we explicitly neglect singlet contributions from the vector coupling to closed fermion loops.

The helicity amplitudes for five-parton scattering at tree level~\cite{Hagiwara:1988pp, Berends:1988yn, Falck:1988gc} and four-parton scattering at one loop~\cite{Glover:1996eh,Campbell:1997tv,Bern:1997sc}, necessary for the NLO calculation, are directly implemented within~\texttt{NNLOJET}. 
Six-parton tree-level and five-parton one-loop amplitudes, required for the double-real and the real-virtual corrections respectively, are obtained from \texttt{OpenLoops2}~\cite{Buccioni:2019sur,Buccioni:2017yxi}, ensuring efficient and stable numerical evaluations across the whole phase space. 

In order to achieve the cancellation of infrared singularities arising in intermediate steps of higher-order calculations, we adopt the antenna subtraction method~\cite{Gehrmann-DeRidder:2005btv,Currie:2013vh}. This entails using \textit{antenna functions} for the construction of counterterms to remove the divergent behavior of real-emission matrix elements in soft and collinear limits. After an inclusive integration over the radiation phase space, which is performed analytically~\cite{Gehrmann-DeRidder:2003pne}, these counterterms are then combined with virtual corrections, subtracting their explicit infrared singularities. 

The assembly of infrared counterterms has typically been a bottleneck in the calculation of higher-order corrections to processes with high partonic multiplicity, such as four-jet production at electron-positron colliders. To address this task in an automated fashion, we rely on the \textit{colorful antenna subtraction method}~\cite{Chen:2022ktf,Gehrmann:2023dxm}, which allows to infer real-emission counterterms from the universal structure of infrared singularities appearing in virtual amplitudes by means of suitable infrared insertion operators in color space. Moreover, we employ \textit{generalized antenna functions}~\cite{Fox:2024bfp} for final-state radiation constructed directly from a target set of unresolved limits, following the algorithmic procedure outlined in Refs.~\cite{Braun-White:2023sgd,Braun-White:2023zwd}. As discussed in Ref.~\cite{Fox:2024bfp}, these offer significant advantages in terms of both simplicity and numerical performance of the infrared subtraction counterterms.

The infrared counterterms are extensively validated by computing their ratio to the respective matrix elements for phase-space points lying increasingly close to soft and collinear configurations, and by observing a gradual improvement in the cancellation of the divergent behavior. Moreover, we verified analytically that all explicit infrared singularities cancel exactly. We validated our setup at lower perturbative orders against \texttt{SHERPA}~\cite{Sherpa:2019gpd,Sherpa:2024mfk}. In particular, we computed the cross section for four- and five-jet production up to NLO and for six-jet production up to LO for different values of the jet resolution parameter $\ycut$, finding excellent agreement. In general, we observe that our implementation up to NNLO, including matrix elements and infrared counterterms, exhibits very good numerical stability, even for low values of $\ycut$ which probe events close to the three-jet (planar) limit. 

The two-loop contributions for this process are novel and state-of-the-art, and we discuss them in more detail below.

\subsection{Two-loop corrections}

The two-loop corrections are implemented in the \texttt{NNLOJET} framework through the UV- and IR-subtracted two-loop amplitudes denoted as \emph{finite remainders}.
Their analytic representation can be written schematically as
\begin{equation} \label{eq:finite-remainder}
  R^{\vec{a}}(\vec{p}) = \sum_{i,j} \; r_{i}(\vec{p}) ~ M^{\vec{a}}_{i j} ~ f_j(\vec{p}) \,,
\end{equation}
where $\vec{p}$ denotes the kinematic variables, and the multi-index $\vec{a}$ labels collectively helicity, color, and coupling information.
The sets $\{ f_j \}$ and $\{r_i\}$ are bases of transcendental special functions and rational coefficients correspondingly,
while $M^{\vec{a}}_{i j}$ are matrices of rational numbers.

A basis of special functions to represent the relevant two-loop five-particle Feynman integrals, dubbed \textit{one-mass pentagon functions} and denoted by $\{f_j^{\text{prod}}\}$, were previously available only in the kinematical crossing corresponding to the hadronic production channel of one massive and two massless particles~\cite{Chicherin:2021dyp,Abreu:2023rco}.
Following the same approach, in this work we construct the complete basis of one-mass pentagon functions for the decay of an off-shell particle into four massless ones, $\{f_j^{\text{decay}}\}$, and implement their efficient evaluation in the public library \texttt{PentagonFunctions++}~\cite{PentagonFunctions}. 
 
We discuss this construction, as well as a number of intriguing observations regarding the functions' divergence structure, in the Appendix.
We stress that, while the two-loop contributions considered in this work only include planar diagrams in the LCA, the set of functions that we present will allow one to include all subleading-color contributions in the future.

Compact analytic expressions for the two-loop finite remainders for the production of an electroweak vector boson in association with two light jets in proton collisions are available for planar contributions~\cite{Abreu:2021asb,DeLaurentis:2025dxw}, which correspond to the approximations discussed in the previous section. The finite remainders are defined in Catani's scheme~\cite{Catani:1998bh,Sterman:2002qn,Becher:2009cu}, which we also employ here. 
The results are written as in Eq.~\eqref{eq:finite-remainder},
with the production-channel functions $\{f_j^{\text{prod}}\}$.
We cross these expressions into the decay channel by replacing the  production pentagon functions with the decay ones (see Appendix). This yields new sets of rational matrices in Eq.~\eqref{eq:finite-remainder} for the finite remainders in the decay channel.
The rational bases $r_i$ from Ref.~\cite{DeLaurentis:2025dxw} are crossed by simply evaluating their expressions with suitable momentum permutations.

Finally, we derive all momentum permutations of Eq.~\eqref{eq:finite-remainder} required to obtain the color- and helicity-summed squared finite remainders entering the double-virtual contributions to $e^+ e^- \to \gamma^*/Z \to jjjj$.
We note that the interference effects in identical-quark channels are also explicitly excluded in the LCA.
We make the implementation available in a public library~\cite{FivePointAmplitudes}, which is interfaced with \texttt{NNLOJET}.
We observe excellent numerical performance of our implementation, similar to the one reported in Ref.~\cite{DeLaurentis:2025dxw} for the production channel, 
which allows us to integrate the two-loop contributions with small computational cost compared to the rest of the calculation.

We validated our implementation by  checking that the renormalization-scale dependence of the finite remainder agrees with the prediction (see \textit{e.g.}\ Ref.~\cite{Abreu:2021oya}),
and that the assembly of the squared matrix elements agrees with the internal implementation in \texttt{NNLOJET} at one loop.

\begin{figure}[t]
\includegraphics[width = 0.5\textwidth]{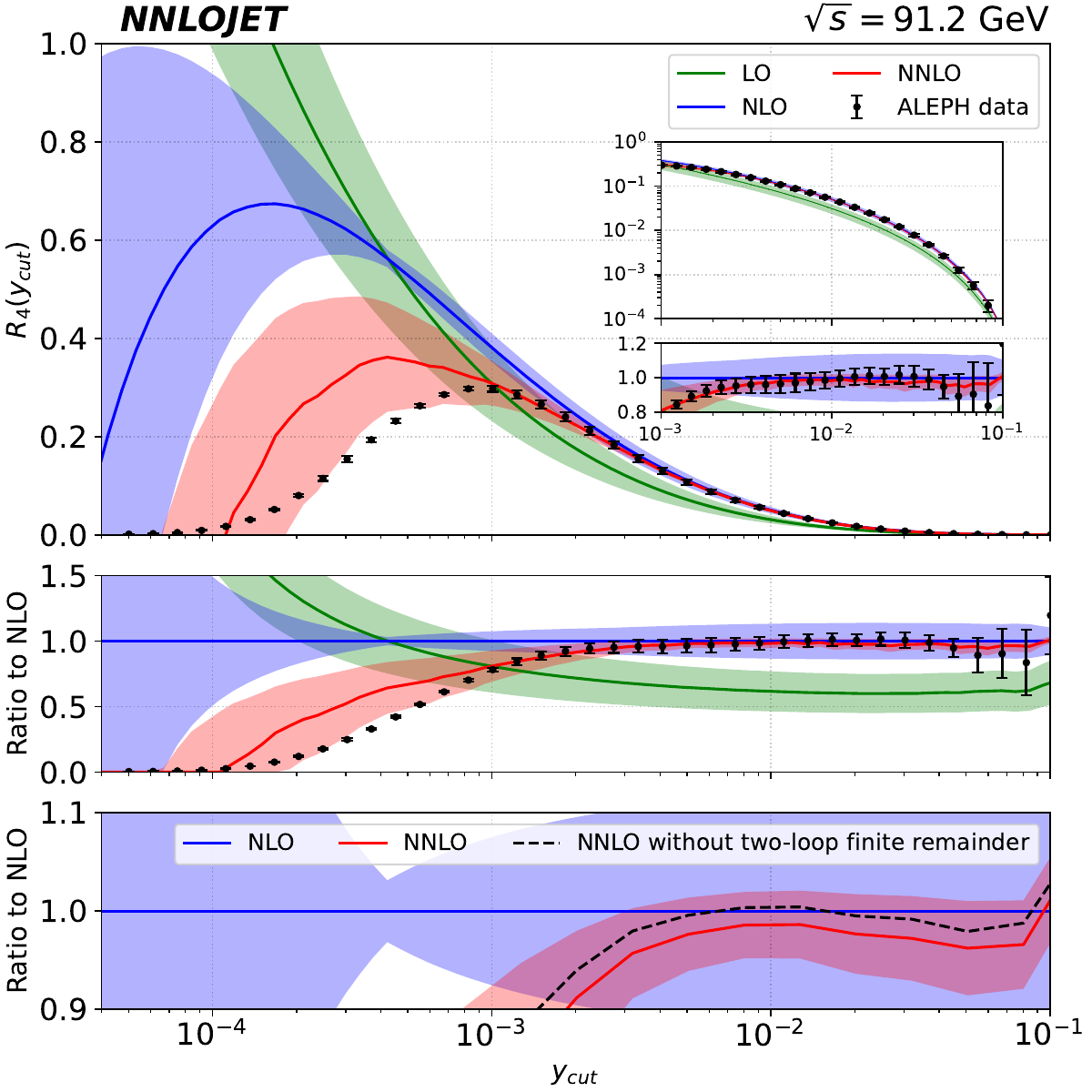}
\caption{Fixed-order predictions for the four-jet rate compared to data from the ALEPH experiment~\cite{ALEPH:2003obs} (black) at \mbox{$\sqrt{s}=91.2$} GeV. Predictions are shown at LO (green), NLO (blue) and NNLO (red). In the inserted frame, we zoom in the $\ycut\in[10^{-3},10^{-1}]$ region and use a logarithmic scale on the $y$-axis. Colored bands represent theory uncertainties obtained by varying the renormalization scale. The ratio to NLO-accurate predictions is shown in the middle frame. In the bottom frame, NNLO results without the two-loop finite remainder (dashed black line) are compared with the complete NNLO predictions (red), both normalized to NLO results (blue).}
\label{fig:4jetRate}
\end{figure}

\section{Results}

We consider $m_Z=91.2$~GeV and \mbox{$\alpha_s(m_Z)=0.118$}. In the denominator of Eq.~\eqref{eq:jet_rate}, the inclusive cross section $\sigma_{\text{inc.}}$ is computed up to the $\alpha_s$ order of the numerator, namely up to $\alpha_s^2$ at LO, $\alpha_s^3$~\cite{Gorishnii:1990vf,Surguladze:1990tg,Chetyrkin:1996ela} at NLO and $\alpha_s^4$~\cite{Baikov:2008jh,Baikov:2010je,Baikov:2012er} at NNLO. This normalization choice ensures that different jet-multiplicity rates add up to $1$ at any order in $\alpha_s$. The renormalization scale is set to $\mu_r=\sqrt{s}$, and the theory uncertainties are estimated by varying $\mu_r$ in the range $\mu_r\in[\sqrt{s}/2,2\sqrt{s}]$ in a correlated manner between the numerator and the denominator of Eq.~\eqref{eq:jet_rate}. The uncertainty bands obtained in this way are then symmetrized~\cite{Cacciari:2011ze}: the maximum variation among the upward and downward ones is assigned symmetrically from the central value. This procedure yields a more conservative estimate of theory uncertainties, as we elaborate in the following. We provide our results in digital format in ancillary files~\cite{zenodo}.

For $\ycut\gtrsim 10^{-1}$, the reconstruction of four distinct jets is unlikely due to the largeness of the resolution parameter, meaning that $R_4(\ycut)$ is very suppressed and no precise data is available in this region. Hence, we set \mbox{$\ycut= 10^{-1}$} as the upper bound of the figures presented in the following. On the other hand, as $\ycut\to 0$, the small jet resolution parameter leads to less inclusive clustering, thereby enhancing the sensitivity to soft and collinear emissions. In any fixed-order calculation, this manifests as the onset of large logarithmic contributions scaling as $\alphas^n\ln^{m}(\ycut)$ with $m\leq 2n$, which must be resummed to all orders in $\alphas$ to restore predictive power. Moreover, other effects like quark masses and hadronization also become important in this region. There is no sharply defined value of $\ycut$  at which fixed-order calculations cease to be valid. Nevertheless, we identify a ``perturbative'' region, $10^{-3} \lesssim  \ycut \lesssim 10^{-1}$, within which fixed-order predictions are expected to provide a reliable description of QCD dynamics and to be systematically improvable through the inclusion of higher-order corrections.

In Fig.~\ref{fig:4jetRate}, we present fixed-order predictions from LO to NNLO for the four-jet rate and compare them to data from the ALEPH experiment~\cite{ALEPH:2003obs} at $\sqrt{s}=91.2$ GeV, where the most accurate data is available. We observe remarkable convergence in the perturbative region, with the NNLO correction being particularly small in the range $3\cdot10^{-3} \lesssim \ycut \lesssim 5\cdot10^{-2}$, corresponding to a negative $2\%$~--~$5\%$ of the full NNLO results. Nevertheless, the shape of the NNLO prediction agrees better with the shape of the data in the perturbative region. Moreover, the theory uncertainties exhibit a significant reduction in size, moving from $\pm15\%$ at NLO to $\pm(3\%-5\%)$ at NNLO.  We note that the conventional scale variation procedure without symmetrization yields, in the perturbative region, an asymmetric theoretical uncertainty of $(+0\%,-(3-5)\%)$, due to $\mu_r=\sqrt{s}$ being close to the maximum of the scale dependence at NNLO, and hence resulting in potentially underestimated uncertainties.
We observe that the residual theory uncertainties at NNLO are smaller than the experimental ones for LEP data, which are dominated by systematic errors. More accurate data, either obtained by reanalyzing existing measurements or collected at future electron-positron colliders, will then enable further high-precision tests of QCD. 

As we move to smaller $\ycut$ values, NNLO corrections grow in size and are negative, leading to a visible deviation from the NLO prediction. In particular, the four-jet rate peak reduces and moves to a higher value of $\ycut$. This brings the NNLO result closer to the experimental data everywhere, especially to the left end of the perturbative region, where the agreement is excellent and significantly improved with respect to the previous order. For $\ycut\lesssim 10^{-3}$, we observe that, while the NNLO predictions still lie closer to the data, they clearly fail to reproduce the measurements, and signs of perturbative breakdown are evident. Specifically, the size of the theory uncertainties dramatically increases, they no longer overlap with the NLO scale variation band, and predictions become unphysical as the NNLO curve turns negative. As explained above, all-order resummation is required in this region. Nevertheless, it is evident that capturing the logarithmically-enhanced terms present in the NNLO correction alone significantly improves the theory prediction.   

In the bottom frame of Fig.~\ref{fig:4jetRate}, we investigate the numerical impact of the two-loop finite remainder in the LCA. In the perturbative region, particularly around $\ycut\approx10^{-2}$, we observe that by omitting the two-loop finite remainder the NNLO correction essentially vanishes. Here, the two-loop finite remainder then represents the entirety of the NNLO correction, and amounts to a negative contribution of $2\%$~--~$5\%$ at the level of the complete NNLO prediction. This pattern is different from what observed in hadronic processes, where the two-loop finite remainder typically represents a smaller contribution to the NNLO correction. Nevertheless its overall impact is similar to what is seen in hadronic processes~\cite{Badger:2023mgf,Abreu:2023bdp,Buccioni:2025bkl,Czakon:2025wgs}.  
Assuming that subleading-color contributions to the two-loop finite remainder are suppressed by a factor of $1/N_c^2\approx 0.1$, we can safely estimate that the uncertainty associated with neglecting them is at most percent level of the full NNLO prediction in the perturbative region, below the size of the current theory uncertainties at NNLO.

\begin{figure}[t]
\includegraphics[width = 0.5\textwidth]{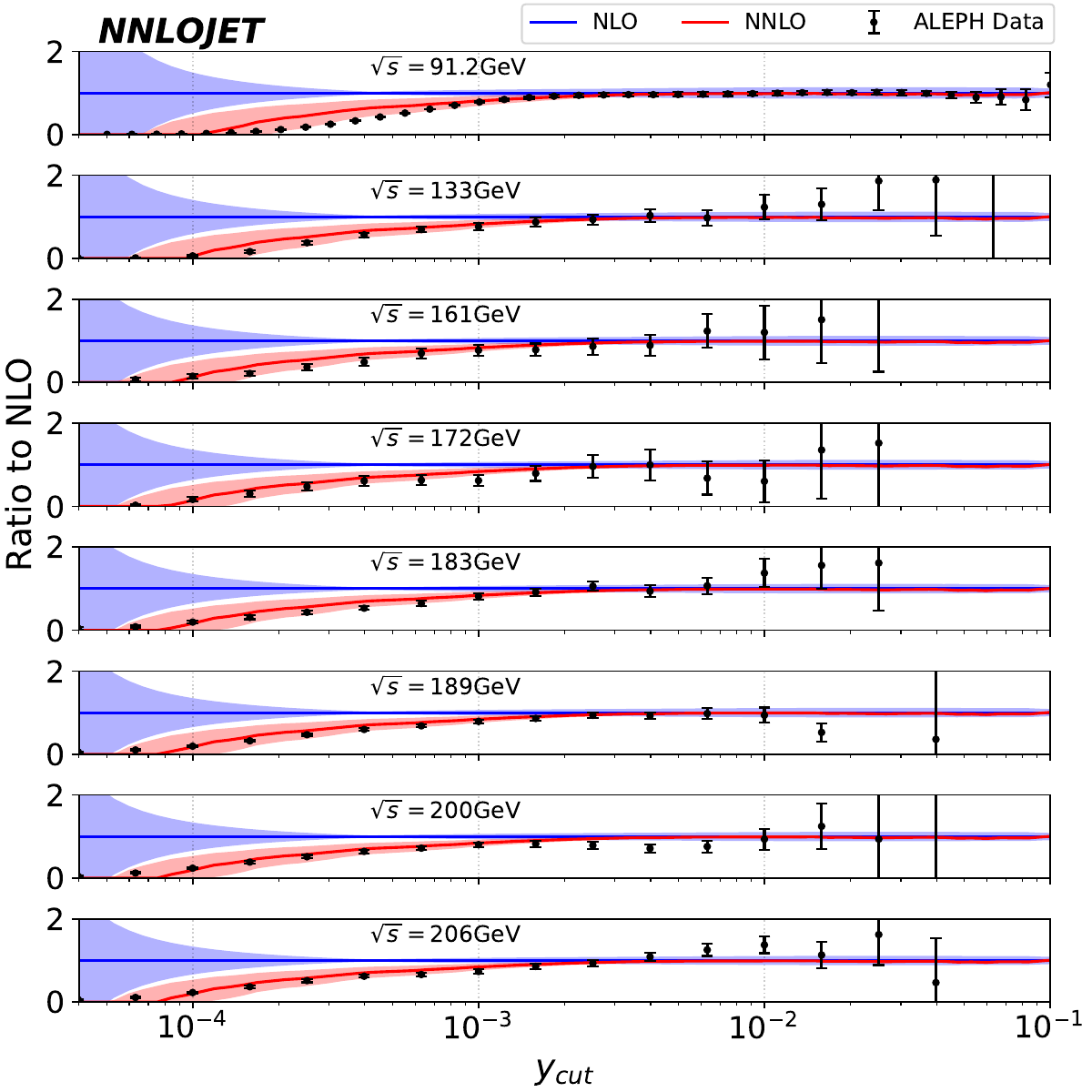}
\caption{Fixed-order predictions for the four-jet rate compared to data from the ALEPH experiment~\cite{ALEPH:2003obs} (black) at several center-of-mass energies. Predictions are shown at NLO (blue) and NNLO (red). Colored bands represent theory uncertainties obtained by varying the renormalization scale.}
\label{fig:4jetRateEnergies}
\end{figure}

In Fig.~\ref{fig:4jetRateEnergies}, we show NLO- and NNLO-accurate predictions compared with ALEPH data at different center-of-mass energies. For a better visualization, we provide results in the same format of Fig.~\ref{fig:4jetRate} for each value of $\sqrt{s}$ as ancillary files~\cite{zenodo}. In general, we observe very similar features to the ones discussed for $\sqrt{s}=91.2$ GeV, namely  very good convergence with small NNLO corrections in the perturbative region, and a sizeable negative shift for $\ycut\lesssim10^{-3}$ which brings the prediction closer to data with respect to the NLO results. At higher energies, the strong coupling is smaller and perturbative calculations should exhibit improved convergence. This also means that the breakdown of fixed-order predictions is pushed to lower values of $\ycut$, and partially explains why the theory calculation agrees better with experimental data for smaller $\ycut$ as the center-of-mass energy increases. Nevertheless, all-order resummation is still necessary to cure the indicators of perturbative breakdown, such as non-overlapping scale variation bands and the NNLO prediction turning negative, as can be seen in Fig.~\ref{fig:4jetRateEnergies}. Regarding the size of the two-loop finite remainder, we observe that it is largely insensitive to the center-of-mass energy (see ancillary files~\cite{zenodo}) and consistently below the theory uncertainties.

\section{Conclusions and Outlook}

In this Letter, we report on the first calculation of the NNLO correction to four-jet production in electron-positron annihilation, and present phenomenological predictions for the four-jet rate. We compare our results with measurements performed by the ALEPH experiment~\cite{ALEPH:2003obs} at LEP, finding improved agreement with data. 

Our calculation represents an important milestone in the advancement of precision QCD phenomenology, especially in view of future electron-positron machines such as FCC-ee and CEPC. 
Future directions include the computation of the  NNLO correction to four-jet event shapes and other observables that probe the deviation from the planar configuration, such as thrust minor~\cite{MARK-J:1980tvz}, D-parameter~\cite{Ellis:1980wv} and aplanarity~\cite{PLUTO:1978jrw}. To correctly describe the region close to the three-jet limit, the all-order resummation of large logarithmic contributions is required. Implementing resummed predictions and matching them to fixed-order results will be the subject of future work. 

\section{Acknowledgments}
\begin{acknowledgments}
The authors would like to thank Yuesheng Dai, Aude Gehrmann-De Ridder, Thomas Gehrmann, Giovanni Stagnitto, Christian Preuss and Max Zoller for discussions.
X.C.\ is supported by the National Science Foundation of China (NSFC) with grants No.~12475085 and No.~12321005. 
D.C.\ is supported by ANR-24-CE31-7996.
N.G.\ is supported by the UK Science and Technology Facilities Council (STFC) through the contract ST/X000745/1.
M.M.\ is supported by a Royal Society Newton International Fellowship (NIF/R1/232539).
V.S.\ has received funding from the European Research Council (ERC) under the European Union’s Horizon 2020 research and innovation programme grant agreement~101019620 (ERC Advanced Grant TOPUP). 
H.Z.\ is supported by the European Union under the Marie Sk{\l}odowska-Curie Actions (MSCA) grant No.~101202083. 
S.Z.\ is supported by the Swiss National Science Foundation (SNSF) under the Ambizione grant No.~215960.
\end{acknowledgments}


\appendix

\section{One-mass pentagon functions in the decay channel}
\label{sec:pentagon-functions}

A complete basis of special functions sufficient to represent any one- and two-loop five-particle finite remainder with a single external massive particle was constructed in Refs.~\cite{Chicherin:2021dyp,Abreu:2023rco}, together with an efficient numerical implementation.
These basis functions, known as (one-mass) pentagon functions, are of polylogarithmic type, i.e.~they can be represented as iterated integrals of logarithmic one-forms~\cite{Abreu:2022mfk}.
The construction in Refs.~\cite{Chicherin:2021dyp,Abreu:2023rco} is however performed in the production channel with two incoming massless particles.
On the one hand, the analytic representation of the basis elements used for numerical evaluation is valid only in that region.
On the other, the basis elements themselves are chosen to minimize the occurrence of spurious singularities --- singularities of the integration kernels where the functions themselves remain finite --- inside the production region.

In this work, we construct a dedicated set of pentagon functions for the decay channel, where all four massless particles are in the final state.
The latter is defined by the following inequalities:
\begin{align}
  p_1^2 >0 \,, \quad
  p_1 \cdot p_i < 0  \,, \quad
  p_i \cdot p_j > 0  \,, \quad \Delta_5 <0 \,,
\end{align}
for $2\le i < j \le 5$, where all momenta are outgoing, $p_1$ is the momentum of the decaying particle, and $\Delta_5$ is the determinant of the Gram matrix of $p_1$, $p_2$, $p_3$ and $p_4$. 

Following Refs.~\cite{Chicherin:2021dyp,Abreu:2023rco}, we solve the canonical differential equations (CDEs)~\cite{Henn:2013pwa} for the two-loop five-particle Feynman integrals obtained in Refs.~\cite{Abreu:2020jxa,Abreu:2021smk,Abreu:2023rco} in terms of iterated integrals.
From the space of solutions, we construct a graded basis of functions and express the basis elements as one-fold path integrals over products of logarithms, dilogarithms, and logarithmic one-forms~\cite{Caron-Huot:2014lda}, evaluated along integration paths lying entirely within the decay region.
This construction relies on the values of the Feynman integrals at an arbitrary base point in the decay channel. The requirement that the symbol-level basis lifts to a function basis~\cite{Abreu:2023rco} imposes strong constraints, fixing most of these values. The remaining constants are determined through numerical evaluations with \texttt{AMFlow}~\cite{Liu:2017jxz,Liu:2022chg}, which also serves as an independent consistency check.
We further crosscheck our results against an independent evaluation of all Feynman integrals at a random point in the decay channel obtained with either \texttt{AMFlow} or \texttt{DiffExp}~\cite{Hidding:2020ytt}.

The arguments of logarithms and dilogarithms are drawn from the alphabet letters appearing in the CDEs~\cite{Duhr:2011zq}. Although the resulting one-fold integrals are well defined in the decay region, the integration paths may cross hypersurfaces where a letter vanishes, requiring an additional analytic treatment of the integrand prior to numerical evaluation~\cite{Chicherin:2021dyp}.
In this respect, the decay channel is simpler than the production channel: only letters that depend quadratically on the Mandelstam invariants can change sign in the production channel.
In particular, the degree-4 polynomial
\begin{align}
\begin{aligned}
& \Sigma_5 = (s_{12} s_{15} - s_{12} s_{23} - s_{15} s_{45} + s_{34} s_{45} + s_{23} s_{34})^2 \\
& \phantom{\Sigma_5 = } - 4 s_{23} s_{34}s_{45}(s_{34} - s_{12} - s_{15})\,,
\end{aligned}
\end{align}
and its permutations, which in Ref.~\cite{Abreu:2023rco} were shown to give rise to divergences within the production channel, have a definite sign throughout the decay channel.
All singularities in the decay channel are thus spurious, and the numerical implementation  follows straightforwardly the approach of Ref.~\cite{Chicherin:2020oor}, leading to improved performance and stability.

The integration paths originate from the base point used in solving the CDEs.
We choose it to be symmetric under permutations of the final-state momenta.
This symmetry reduces the number of values needed to solve the CDEs and simplifies the iterated-integral representation of the solution by reducing the proliferation of distinct transcendental constants.
In contrast to the production channel, the permutation symmetry constrains the base point up to its normalization;
we choose it to be
\begin{align} \label{eq:X_0}
  p_1^2 = 1 \,, \quad \ s_{12} = s_{15} = \frac12 \,, \quad \ s_{23} = s_{34} = s_{45} = \frac16 \,,
\end{align} 
where $s_{ij} = (p_i+p_j)^2$.
Interestingly, all letters that can vanish within the decay channel do so at the base point.

The endpoint of the integration path, on the other hand, specifies the kinematic point at which the pentagon functions are evaluated.
As the production channel, also the decay one --- expressed in Mandelstam variables --- is not star-shaped: straight paths originating from the base point in Eq.~\eqref{eq:X_0} are not always sufficient to reach any point in the decay channel without leaving it.
In practice, we observe that two straight segments lying within the decay region suffice, with an intermediate point chosen according to the Monte Carlo strategy of Ref.~\cite{Chicherin:2021dyp}.

We make the numerical evaluation of the decay-channel pentagon functions publicly available through the \textsc{C++} library \texttt{PentagonFunctions++}~\cite{PentagonFunctions}.
Efficiency and stability of this evaluation routine are certified by the phenomenological study presented in this Letter.
Symbolic expressions of all relevant master integrals in terms of decay-channel pentagon functions, and expressions of the latter in terms of either iterated or one-fold integrals can be found in ancillary files~\cite{zenodo}.
Additionally, we provide the mapping between the production- and decay-channel pentagon functions.
Crossing from the production to the decay channel then simply amounts to replacing each production-channel pentagon function with the corresponding polynomial in the decay-channel pentagon functions.
Finally, we stress that these results include also the non-planar pentagon functions that do not appear in the LCA employed in this work.


\bibliography{main}

@article{Duhr:2011zq,
    author = "Duhr, Claude and Gangl, Herbert and Rhodes, John R.",
    title = "{From polygons and symbols to polylogarithmic functions}",
    eprint = "1110.0458",
    archivePrefix = "arXiv",
    primaryClass = "math-ph",
    reportNumber = "IPPP-11-56, DCPT-11-112",
    doi = "10.1007/JHEP10(2012)075",
    journal = "JHEP",
    volume = "10",
    pages = "075",
    year = "2012"
}

@software{zenodo,
  author        = {Chen, Xuan and Chicherin, Dmitry and Fox, Elliot and Glover, Nigel and Marcoli, Matteo and Sotnikov, Vasily and Sun, Huiting and Zhang, Hantian and Zoia, Simone},
  title         = {{Supplementary material for ``The Four-Jet Rate in Electron-Positron Annihilation at Order $\alpha_s^4$}},
  publisher     = {Zenodo},
  year          = "2026",
  month         = "2",
  note = "{DOI} \href{https://doi.org/10.5281/zenodo.18630975}{10.5281/zenodo.18630975}"
}

@article{Abreu:2022mfk,
    author = "Abreu, Samuel and Britto, Ruth and Duhr, Claude",
    title = "{The SAGEX review on scattering amplitudes Chapter 3: Mathematical structures in Feynman integrals}",
    eprint = "2203.13014",
    archivePrefix = "arXiv",
    primaryClass = "hep-th",
    reportNumber = "SAGEX-22-04, BONN-TH-2022-03, CERN-TH-2022-021",
    doi = "10.1088/1751-8121/ac87de",
    journal = "J. Phys. A",
    volume = "55",
    number = "44",
    pages = "443004",
    year = "2022"
}

@article{Hidding:2020ytt,
    author = "Hidding, Martijn",
    title = "{DiffExp, a Mathematica package for computing Feynman integrals in terms of one-dimensional series expansions}",
    eprint = "2006.05510",
    archivePrefix = "arXiv",
    primaryClass = "hep-ph",
    doi = "10.1016/j.cpc.2021.108125",
    journal = "Comput. Phys. Commun.",
    volume = "269",
    pages = "108125",
    year = "2021"
}

@article{Liu:2017jxz,
    author = "Liu, Xiao and Ma, Yan-Qing and Wang, Chen-Yu",
    title = "{A Systematic and Efficient Method to Compute Multi-loop Master Integrals}",
    eprint = "1711.09572",
    archivePrefix = "arXiv",
    primaryClass = "hep-ph",
    doi = "10.1016/j.physletb.2018.02.026",
    journal = "Phys. Lett. B",
    volume = "779",
    pages = "353--357",
    year = "2018"
}

@article{Henn:2013pwa,
    author = "Henn, Johannes M.",
    title = "{Multiloop integrals in dimensional regularization made simple}",
    eprint = "1304.1806",
    archivePrefix = "arXiv",
    primaryClass = "hep-th",
    doi = "10.1103/PhysRevLett.110.251601",
    journal = "Phys. Rev. Lett.",
    volume = "110",
    pages = "251601",
    year = "2013"
}

@article{Caron-Huot:2014lda,
    author = "Caron-Huot, Simon and Henn, Johannes M.",
    title = "{Iterative structure of finite loop integrals}",
    eprint = "1404.2922",
    archivePrefix = "arXiv",
    primaryClass = "hep-th",
    doi = "10.1007/JHEP06(2014)114",
    journal = "JHEP",
    volume = "06",
    pages = "114",
    year = "2014"
}

@article{Abreu:2023rco,
    author = "Abreu, Samuel and Chicherin, Dmitry and Ita, Harald and Page, Ben and Sotnikov, Vasily and Tschernow, Wladimir and Zoia, Simone",
    title = "{All Two-Loop Feynman Integrals for Five-Point One-Mass Scattering}",
    eprint = "2306.15431",
    archivePrefix = "arXiv",
    primaryClass = "hep-ph",
    reportNumber = "CERN-TH-2023-119",
    doi = "10.1103/PhysRevLett.132.141601",
    journal = "Phys. Rev. Lett.",
    volume = "132",
    number = "14",
    pages = "141601",
    year = "2024"
}

@article{Chicherin:2020oor,
    author = "Chicherin, Dmitry and Sotnikov, Vasily",
    title = "{Pentagon Functions for Scattering of Five Massless Particles}",
    eprint = "2009.07803",
    archivePrefix = "arXiv",
    primaryClass = "hep-ph",
    reportNumber = "MPP-2020-171",
    doi = "10.1007/JHEP12(2020)167",
    journal = "JHEP",
    volume = "20",
    pages = "167",
    year = "2020"
}

@article{Chicherin:2021dyp,
    author = "Chicherin, Dmitry and Sotnikov, Vasily and Zoia, Simone",
    title = "{Pentagon functions for one-mass planar scattering amplitudes}",
    eprint = "2110.10111",
    archivePrefix = "arXiv",
    primaryClass = "hep-ph",
    reportNumber = "LAPTH-041/21, MPP-2021-182",
    doi = "10.1007/JHEP01(2022)096",
    journal = "JHEP",
    volume = "01",
    pages = "096",
    year = "2022"
}

@article{Abreu:2020jxa,
  title = {Two-{{Loop Integrals}} for {{Planar Five-Point One-Mass Processes}}},
  author = {Abreu, Samuel and Ita, Harald and Moriello, Francesco and Page, Ben and Tschernow, Wladimir and Zeng, Mao},
  year = 2020,
  journal = {JHEP},
  volume = {11},
  eprint = {2005.04195},
  primaryclass = {hep-ph},
  pages = {117},
  doi = {10.1007/JHEP11(2020)117},
  archiveprefix = {arXiv}
}

@article{Abreu:2021smk,
  title = {Two-Loop Hexa-Box Integrals for Non-Planar Five-Point One-Mass Processes},
  author = {Abreu, Samuel and Ita, Harald and Page, Ben and Tschernow, Wladimir},
  year = 2022,
  month = mar,
  journal = {JHEP},
  volume = {03},
  eprint = {2107.14180},
  primaryclass = {hep-ph},
  pages = {182},
  doi = {10.1007/JHEP03(2022)182},
  archiveprefix = {arXiv}
}

@article{DeLaurentis:2025dxw,
    author = "De Laurentis, Giuseppe and Ita, Harald and Page, Ben and Sotnikov, Vasily",
    title = "{Compact two-loop QCD corrections for Vjj production in proton collisions}",
    eprint = "2503.10595",
    archivePrefix = "arXiv",
    primaryClass = "hep-ph",
    reportNumber = "PSI-PR-24-29, ZU-TH 14/25",
    doi = "10.1007/JHEP06(2025)093",
    journal = "JHEP",
    volume = "06",
    pages = "093",
    year = "2025"
}

@article{Dixon:1997th,
  title = {Complete {{O}}($\alpha_s^3$) Results for $e^+ e^- \to (\gamma, {{Z}}) \to$ Four Jets},
  author = {Dixon, Lance J and Signer, Adrian},
  year = 1997,
  journal = {Phys. Rev. D},
  volume = {56},
  eprint = {hep-ph/9706285},
  pages = {4031--4038},
  doi = {10.1103/PhysRevD.56.4031},
  archiveprefix = {arXiv}
}

@article{Kniehl:1989bb,
  title = {{{QCD Corrections}} to the {{Axial Part}} of the {{Z Decay Rate}}},
  author = {Kniehl, Bernd A and Kuhn, Johann H},
  year = 1989,
  journal = {Phys. Lett. B},
  volume = {224},
  pages = {229--232},
  issn = {0370-2693},
  doi = {10.1016/0370-2693(89)91079-4}
}

@article{Signer:1996bf,
  title = {Electron - Positron Annihilation into Four Jets at next-to-Leading Order in $\alpha_s$},
  author = {Signer, Adrian and Dixon, Lance J},
  year = 1997,
  journal = {Phys. Rev. Lett.},
  volume = {78},
  eprint = {hep-ph/9609460},
  pages = {811--814},
  issn = {0031-9007},
  doi = {10.1103/PhysRevLett.78.811},
  archiveprefix = {arXiv}
}

@article{Czakon:2025wgs,
    author = "Czakon, Michal and Poncelet, Rene",
    title = "{How much color do we really need? Two-loop subleading-color effects in photon and jet physics}",
    eprint = "2512.17591",
    archivePrefix = "arXiv",
    primaryClass = "hep-ph",
    reportNumber = "IFJPAN-IV-2025-24, P3H-25-115, TTK-25-46",
    month = "12",
    year = "2025",
    journal=""
}

@article{Abreu:2023bdp,
  title = {Two-Loop {{QCD}} Corrections for Three-Photon Production at Hadron Colliders},
  author = {Abreu, Samuel and De Laurentis, Giuseppe and Ita, Harald and Klinkert, Maximillian and Page, Ben and Sotnikov, Vasily},
  year = 2023,
  month = oct,
  journal = {SciPost Physics},
  volume = {15},
  number = {4},
  eprint = {2305.17056},
  primaryclass = {hep-ph},
  pages = {157},
  issn = {2542-4653},
  doi = {10.21468/SciPostPhys.15.4.157},
  urldate = {2024-02-05},
  archiveprefix = {arXiv}
}

@article{Badger:2023mgf,
  title = {Isolated Photon Production in Association with a Jet Pair through Next-to-next-to-Leading Order in {{QCD}}},
  author = {Badger, Simon and Czakon, Michal and Hartanto, Heribertus Bayu and Moodie, Ryan and Peraro, Tiziano and Poncelet, Rene and Zoia, Simone},
  year = 2023,
  month = oct,
  journal = {arXiv [hep-ph]},
  volume = {10},
  eprint = {2304.06682},
  primaryclass = {hep-ph},
  pages = {071},
  doi = {10.1007/JHEP10(2023)071},
  urldate = {2023-04-14},
  archiveprefix = {arXiv},
  isbn = {2304.06682}
}

@article{Liu:2022chg,
  title = {{{AMFlow}}: A {{Mathematica}} Package for {{Feynman}} Integrals Computation via {{Auxiliary Mass Flow}}},
  author = {Liu, Xiao and Ma, Yan-Qing},
  year = 2023,
  month = feb,
  journal = {Comput. Phys. Commun.},
  volume = {283},
  eprint = {2201.11669},
  primaryclass = {hep-ph},
  pages = {108565},
  doi = {10.1016/j.cpc.2022.108565},
  archiveprefix = {arXiv},
  isbn = {2201.11669}
}

@article{NNLOJET:2025rno,
    author = "Huss, A. and others",
    collaboration = "NNLOJET",
    title = "{NNLOJET: a parton-level event generator for jet cross sections at NNLO QCD accuracy}",
    eprint = "2503.22804",
    archivePrefix = "arXiv",
    primaryClass = "hep-ph",
    reportNumber = "CERN-TH-2025-012, IPPP/25/09, ZU-TH 11/25",
    month = "3",
    year = "2025",
    journal = ""
}

@article{Hagiwara:1988pp,
    author = "Hagiwara, Kaoru and Zeppenfeld, D.",
    title = "{Amplitudes for Multiparton Processes Involving a Current at $e^+ e^-$, $e^{\pm} p$, and Hadron Colliders}",
    reportNumber = "KEK-TH-199, MAD-PH-402, KEK-PREPRINT-87-158",
    doi = "10.1016/0550-3213(89)90397-0",
    journal = "Nucl. Phys. B",
    volume = "313",
    pages = "560--594",
    year = "1989"
}

@article{Berends:1988yn,
    author = "Berends, Frits A. and Giele, W. T. and Kuijf, H.",
    title = "{Exact Expressions for Processes Involving a Vector Boson and Up to Five Partons}",
    reportNumber = "Print-89-0055 (LEIDEN)",
    doi = "10.1016/0550-3213(89)90242-3",
    journal = "Nucl. Phys. B",
    volume = "321",
    pages = "39--82",
    year = "1989"
}

@article{Falck:1988gc,
    author = "Falck, N. K. and Graudenz, D. and Kramer, G.",
    title = "{Five Jet Production in $e^+ e^-$ Annihilation}",
    reportNumber = "DESY-88-186",
    doi = "10.1016/0370-2693(89)90056-7",
    journal = "Phys. Lett. B",
    volume = "220",
    pages = "299--302",
    year = "1989"
}

@article{Campbell:1997tv,
    author = "Campbell, John M. and Glover, E. W. Nigel and Miller, D. J.",
    title = "{The One loop QCD corrections for $\gamma^* \to q \bar{q} gg$}",
    eprint = "hep-ph/9706297",
    archivePrefix = "arXiv",
    reportNumber = "DTP-97-44, RAL-TR-97-027",
    doi = "10.1016/S0370-2693(97)00909-X",
    journal = "Phys. Lett. B",
    volume = "409",
    pages = "503--508",
    year = "1997"
}

@article{Glover:1996eh,
    author = "Glover, E. W. Nigel and Miller, D. J.",
    title = "{The One loop QCD corrections for $\gamma^* \to Q\bar{Q}q\bar{q}$}",
    eprint = "hep-ph/9609474",
    archivePrefix = "arXiv",
    reportNumber = "DTP-96-66",
    doi = "10.1016/S0370-2693(97)00113-5",
    journal = "Phys. Lett. B",
    volume = "396",
    pages = "257--263",
    year = "1997"
}

@article{Bern:1997sc,
    author = "Bern, Zvi and Dixon, Lance J. and Kosower, David A.",
    title = "{One loop amplitudes for e$^{+}$e$^{-}$ to four partons}",
    eprint = "hep-ph/9708239",
    archivePrefix = "arXiv",
    reportNumber = "SLAC-PUB-7529, SACLAY-SPH-T-97-090, UCLA-97-TEP-10",
    doi = "10.1016/S0550-3213(97)00703-7",
    journal = "Nucl. Phys. B",
    volume = "513",
    pages = "3--86",
    year = "1998"
}

@article{Buccioni:2019sur,
    author = {Buccioni, Federico and Lang, Jean-Nicolas and Lindert, Jonas M. and Maierh\"ofer, Philipp and Pozzorini, Stefano and Zhang, Hantian and Zoller, Max F.},
    title = "{OpenLoops 2}",
    eprint = "1907.13071",
    archivePrefix = "arXiv",
    primaryClass = "hep-ph",
    reportNumber = "IPPP/19/62, FR-PHENO-2019-12, PSI-PR-19-15, ZU-TH 37/19",
    doi = "10.1140/epjc/s10052-019-7306-2",
    journal = "Eur. Phys. J. C",
    volume = "79",
    number = "10",
    pages = "866",
    year = "2019"
}

@article{Buccioni:2017yxi,
    author = "Buccioni, Federico and Pozzorini, Stefano and Zoller, Max",
    title = "{On-the-fly reduction of open loops}",
    eprint = "1710.11452",
    archivePrefix = "arXiv",
    primaryClass = "hep-ph",
    reportNumber = "ZU-TH-29-17",
    doi = "10.1140/epjc/s10052-018-5562-1",
    journal = "Eur. Phys. J. C",
    volume = "78",
    number = "1",
    pages = "70",
    year = "2018"
}

@article{Gehrmann-DeRidder:2005btv,
	author = "Gehrmann-De Ridder, A. and Gehrmann, T. and Glover, E. W. Nigel",
	title = "{Antenna subtraction at NNLO}",
	eprint = "hep-ph/0505111",
	archivePrefix = "arXiv",
	reportNumber = "ZU-TH-07-05, IPPP-05-18",
	doi = "10.1088/1126-6708/2005/09/056",
	journal = "JHEP",
	volume = "09",
	pages = "056",
	year = "2005"
}

@article{Currie:2013vh,
	author = "Currie, James and Glover, E. W. N. and Wells, Steven",
	title = "{Infrared Structure at NNLO Using Antenna Subtraction}",
	eprint = "1301.4693",
	archivePrefix = "arXiv",
	primaryClass = "hep-ph",
	reportNumber = "IPPP-12-82, ZU-TH-26-12",
	doi = "10.1007/JHEP04(2013)066",
	journal = "JHEP",
	volume = "04",
	pages = "066",
	year = "2013"
}

@article{Fox:2024bfp,
	author = "Fox, Elliot and Glover, Nigel and Marcoli, Matteo",
	title = "{Generalised antenna functions for higher-order calculations}",
	eprint = "2410.12904",
	archivePrefix = "arXiv",
	primaryClass = "hep-ph",
	reportNumber = "IPPP/24/63",
	doi = "10.1007/JHEP12(2024)225",
	journal = "JHEP",
	volume = "12",
	pages = "225",
	year = "2024"
}

@article{Braun-White:2023zwd,
	author = "Braun-White, Oscar and Glover, Nigel and Preuss, Christian T.",
	title = "{A general algorithm to build mixed real and virtual antenna functions for higher-order calculations}",
	eprint = "2307.14999",
	archivePrefix = "arXiv",
	primaryClass = "hep-ph",
	reportNumber = "IPPP/23/31, ZU-TH 30/23",
	doi = "10.1007/JHEP11(2023)179",
	journal = "JHEP",
	volume = "11",
	pages = "179",
	year = "2023"
}

@article{Braun-White:2023sgd,
	author = "Braun-White, Oscar and Glover, Nigel and Preuss, Christian T.",
	title = "{A general algorithm to build real-radiation antenna functions for higher-order calculations}",
	eprint = "2302.12787",
	archivePrefix = "arXiv",
	primaryClass = "hep-ph",
	reportNumber = "IPPP 23/7",
	doi = "10.1007/JHEP06(2023)065",
	journal = "JHEP",
	volume = "06",
	pages = "065",
	year = "2023"
}

@article{Gehrmann-DeRidder:2008qsl,
	author = "Gehrmann-De Ridder, A. and Gehrmann, T. and Glover, E. W. N. and Heinrich, G.",
	title = "{Jet rates in electron-positron annihilation at O$(\alpha_s^3$) in QCD}",
	eprint = "0802.0813",
	archivePrefix = "arXiv",
	primaryClass = "hep-ph",
	reportNumber = "ZU-TH-03-08, IPPP-08-05",
	doi = "10.1103/PhysRevLett.100.172001",
	journal = "Phys. Rev. Lett.",
	volume = "100",
	pages = "172001",
	year = "2008"
}

@article{Gehrmann:2023dxm,
    author = "Gehrmann, T. and Glover, E. W. N. and Marcoli, M.",
    title = "{The colourful antenna subtraction method}",
    eprint = "2310.19757",
    archivePrefix = "arXiv",
    primaryClass = "hep-ph",
    reportNumber = "IPPP/23/63 ZU-TH 70/23",
    doi = "10.1007/JHEP03(2024)114",
    journal = "JHEP",
    volume = "03",
    pages = "114",
    year = "2024"
}

@article{Chen:2022ktf,
    author = "Chen, Xuan and Gehrmann, Thomas and Glover, E. W. N. and Huss, Alexander and Marcoli, Matteo",
    title = "{Automation of antenna subtraction in colour space: gluonic processes}",
    eprint = "2203.13531",
    archivePrefix = "arXiv",
    primaryClass = "hep-ph",
    reportNumber = "CERN-TH-2022-047, IPPP/22/15, KA-TP-05-2022, P3H-22-030, ZU-TH 10/22",
    doi = "10.1007/JHEP10(2022)099",
    journal = "JHEP",
    volume = "10",
    pages = "099",
    year = "2022"
}

@article{Catani:1991hj,
    author = "Catani, S. and Dokshitzer, Yuri L. and Olsson, M. and Turnock, G. and Webber, B. R.",
    title = "{New clustering algorithm for multi - jet cross-sections in $e^+ e^-$ annihilation}",
    reportNumber = "CAVENDISH-HEP-91-5",
    doi = "10.1016/0370-2693(91)90196-W",
    journal = "Phys. Lett. B",
    volume = "269",
    pages = "432--438",
    year = "1991"
}

@article{Brown:1990nm,
    author = "Brown, N. and Stirling, W. James",
    title = "{Jet cross-sections at leading double logarithm in $e^+ e^-$ annihilation}",
    reportNumber = "RAL-90-060, DTP-90-58",
    doi = "10.1016/0370-2693(90)90502-W",
    journal = "Phys. Lett. B",
    volume = "252",
    pages = "657--662",
    year = "1990"
}

@article{Brown:1991hx,
    author = "Brown, N. and Stirling, W. James",
    title = "{Finding jets and summing soft gluons: A New algorithm}",
    reportNumber = "RAL-91-049, DTP-91-30",
    doi = "10.1007/BF01559740",
    journal = "Z. Phys. C",
    volume = "53",
    pages = "629--636",
    year = "1992"
}

@article{Stirling:1991ds,
    author = "Stirling, W. James",
    title = "{Hard QCD working group: Theory summary}",
    doi = "10.1088/0954-3899/17/10/014",
    journal = "J. Phys. G",
    volume = "17",
    pages = "1567--1574",
    year = "1991"
}

@article{Bethke:1991wk,
    author = "Bethke, S. and Kunszt, Z. and Soper, D. E. and Stirling, W. James",
    title = "{New jet cluster algorithms: Next-to-leading order QCD and hadronization corrections}",
    reportNumber = "CERN-TH-6222-91",
    doi = "10.1016/0550-3213(92)90289-N",
    journal = "Nucl. Phys. B",
    volume = "370",
    pages = "310--334",
    year = "1992",
    note = "[Erratum: Nucl.Phys.B 523, 681--681 (1998)]"
}

@article{Frederix:2010ne,
    author = "Frederix, Rikkert and Frixione, Stefano and Melnikov, Kirill and Zanderighi, Giulia",
    title = "{NLO QCD corrections to five-jet production at LEP and the extraction of $\alpha_s(M_Z)$}",
    eprint = "1008.5313",
    archivePrefix = "arXiv",
    primaryClass = "hep-ph",
    reportNumber = "CERN-TH-2010-185, OUTP-1019P, ZU-TH-11-10",
    doi = "10.1007/JHEP11(2010)050",
    journal = "JHEP",
    volume = "11",
    pages = "050",
    year = "2010"
}

@article{Chen:2025kez,
    author = "Chen, Xuan and Jakub{\v{c}}{\'\i}k, Petr and Marcoli, Matteo and Stagnitto, Giovanni",
    title = "{Jet production at electron-positron colliders at next-to-next-to-next-to-leading order in QCD}",
    eprint = "2505.10618",
    archivePrefix = "arXiv",
    primaryClass = "hep-ph",
    reportNumber = "IPPP/25/27, ZU-TH 36/25",
    doi = "10.1016/j.physletb.2025.139804",
    journal = "Phys. Lett. B",
    volume = "869",
    pages = "139804",
    year = "2025"
}

@article{Gehrmann-DeRidder:2014hxk,
    author = "Gehrmann-De Ridder, A. and Gehrmann, T. and Glover, E. W. N. and Heinrich, G.",
    title = "{EERAD3: Event shapes and jet rates in electron-positron annihilation at order $\alpha_s^3$}",
    eprint = "1402.4140",
    archivePrefix = "arXiv",
    primaryClass = "hep-ph",
    reportNumber = "ZU-TH-05-14, IPPP-14-15, MPP-2014-23",
    doi = "10.1016/j.cpc.2014.07.024",
    journal = "Comput. Phys. Commun.",
    volume = "185",
    pages = "3331",
    year = "2014"
}

@article{Chetyrkin:1996ela,
    author = "Chetyrkin, K. G. and Kuhn, Johann H. and Kwiatkowski, A.",
    title = "{QCD corrections to the $e^{+} e^{-}$ cross-section and the $Z$ boson decay rate}",
    eprint = "hep-ph/9503396",
    archivePrefix = "arXiv",
    reportNumber = "LBL-36678-REV, TTP-94-32, LBL-36678, MPI-PHT-96-019",
    doi = "10.1016/S0370-1573(96)00012-9",
    journal = "Phys. Rept.",
    volume = "277",
    pages = "189--281",
    year = "1996"
}

@article{Baikov:2008jh,
    author = "Baikov, P. A. and Chetyrkin, K. G. and Kuhn, Johann H.",
    title = "{Order $\alpha_s^4$ QCD Corrections to Z and tau Decays}",
    eprint = "0801.1821",
    archivePrefix = "arXiv",
    primaryClass = "hep-ph",
    reportNumber = "SFB-CPP-08-04, TTP08-01",
    doi = "10.1103/PhysRevLett.101.012002",
    journal = "Phys. Rev. Lett.",
    volume = "101",
    pages = "012002",
    year = "2008"
}

@article{Baikov:2010je,
    author = "Baikov, P. A. and Chetyrkin, K. G. and Kuhn, J. H.",
    title = "{Adler Function, Bjorken Sum Rule, and the Crewther Relation to Order $\alpha^4_s$ in a General Gauge Theory}",
    eprint = "1001.3606",
    archivePrefix = "arXiv",
    primaryClass = "hep-ph",
    reportNumber = "TTP10-05",
    doi = "10.1103/PhysRevLett.104.132004",
    journal = "Phys. Rev. Lett.",
    volume = "104",
    pages = "132004",
    year = "2010"
}

@article{Baikov:2012er,
    author = "Baikov, P. A. and Chetyrkin, K. G. and Kuhn, J. H. and Rittinger, J.",
    title = "{Complete ${\cal O}(\alpha_s^4)$ QCD Corrections to Hadronic $Z$-Decays}",
    eprint = "1201.5804",
    archivePrefix = "arXiv",
    primaryClass = "hep-ph",
    reportNumber = "SFB-CPP-12-05, TTP11-31",
    doi = "10.1103/PhysRevLett.108.222003",
    journal = "Phys. Rev. Lett.",
    volume = "108",
    pages = "222003",
    year = "2012"
}

@article{Gorishnii:1990vf,
    author = "Gorishnii, S. G. and Kataev, A. L. and Larin, S. A.",
    title = "{The $O(\alpha^{3}_{s})$-corrections to $\sigma_{tot}(e^{+}e^{-}\rightarrow hadrons)$ and $\Gamma(\tau^{-} \rightarrow \nu_{\tau} + hadrons)$ in QCD}",
    reportNumber = "UM-TH-91-01",
    doi = "10.1016/0370-2693(91)90149-K",
    journal = "Phys. Lett. B",
    volume = "259",
    pages = "144--150",
    year = "1991"
}

@article{Surguladze:1990tg,
    author = "Surguladze, Levan R. and Samuel, Mark A.",
    title = "{Total hadronic cross-section in $e^+ e^-$ annihilation at the four loop level of perturbative QCD}",
    reportNumber = "OSU-RN-250",
    doi = "10.1103/PhysRevLett.66.560",
    journal = "Phys. Rev. Lett.",
    volume = "66",
    pages = "560--563",
    year = "1991",
    note = "[Erratum: Phys.Rev.Lett. 66, 2416 (1991)]"
}

@article{Hagiwara:1990dx,
    author = "Hagiwara, Kaoru and Kuruma, T. and Yamada, Y.",
    title = "{Three jet distributions from the one loop Z g g vertex at $e^+ e^-$ colliders}",
    reportNumber = "KEK-TH-265, UT-570-TOKYO, KEK-PREPRINT-90-126",
    doi = "10.1016/0550-3213(91)90532-3",
    journal = "Nucl. Phys. B",
    volume = "358",
    pages = "80--96",
    year = "1991"
}

@article{Weinzierl:2008iv,
    author = "Weinzierl, Stefan",
    title = "{NNLO corrections to 3-jet observables in electron-positron annihilation}",
    eprint = "0807.3241",
    archivePrefix = "arXiv",
    primaryClass = "hep-ph",
    reportNumber = "MZ-TH-08-22",
    doi = "10.1103/PhysRevLett.101.162001",
    journal = "Phys. Rev. Lett.",
    volume = "101",
    pages = "162001",
    year = "2008"
}

@article{Becker:2011vg,
    author = "Becker, Sebastian and Goetz, Daniel and Reuschle, Christian and Schwan, Christopher and Weinzierl, Stefan",
    title = "{NLO results for five, six and seven jets in electron-positron annihilation}",
    eprint = "1111.1733",
    archivePrefix = "arXiv",
    primaryClass = "hep-ph",
    doi = "10.1103/PhysRevLett.108.032005",
    journal = "Phys. Rev. Lett.",
    volume = "108",
    pages = "032005",
    year = "2012"
}

@article{Ellis:1980wv,
    author = "Ellis, R. Keith and Ross, D. A. and Terrano, A. E.",
    title = "{The Perturbative Calculation of Jet Structure in $e^+ e^-$ Annihilation}",
    reportNumber = "CALT-68-785",
    doi = "10.1016/0550-3213(81)90165-6",
    journal = "Nucl. Phys. B",
    volume = "178",
    pages = "421--456",
    year = "1981"
}

@article{Kunszt:1980vt,
    author = "Kunszt, Zoltan",
    title = "{Comment on the O$(\alpha_s^2$) Corrections to Jet Production in $e^+ e^-$ Annihilation}",
    reportNumber = "DESY-80-79",
    doi = "10.1016/0370-2693(81)90563-3",
    journal = "Phys. Lett. B",
    volume = "99",
    pages = "429--432",
    year = "1981"
}

@article{MARK-J:1980tvz,
    author = "Barber, D. P. and others",
    collaboration = "MARK-J",
    title = "{Physics with high-energy electron positron colliding beams with the MARK-J detector}",
    doi = "10.1016/0370-1573(80)90169-6",
    journal = "Phys. Rept.",
    volume = "63",
    pages = "337--391",
    year = "1980"
}

@article{PLUTO:1978jrw,
    author = "Berger, Christoph and others",
    collaboration = "PLUTO",
    title = "{Jet Analysis of the $\Upsilon$ (9.46) Decay Into Charged Hadrons}",
    reportNumber = "DESY-78-71",
    doi = "10.1016/0370-2693(79)90265-X",
    journal = "Phys. Lett. B",
    volume = "82",
    pages = "449--455",
    year = "1979"
}

@article{Vermaseren:1980qz,
    author = "Vermaseren, J. A. M. and Gaemers, K. J. F. and Oldham, S. J.",
    title = "{Perturbative QCD Calculation of Jet Cross-Sections in $e^+ e^-$ Annihilation}",
    reportNumber = "CERN-TH-3002",
    doi = "10.1016/0550-3213(81)90276-5",
    journal = "Nucl. Phys. B",
    volume = "187",
    pages = "301--320",
    year = "1981"
}

@article{Fabricius:1981sx,
    author = "Fabricius, K. and Schmitt, I. and Kramer, G. and Schierholz, G.",
    title = "{Higher Order Perturbative QCD Calculation of Jet Cross-Sections in $e^+ e^-$ Annihilation}",
    reportNumber = "DESY-81-035",
    doi = "10.1007/BF01578281",
    journal = "Z. Phys. C",
    volume = "11",
    pages = "315",
    year = "1981"
}

@article{TASSO:1979zyf,
    author = "Brandelik, R. and others",
    collaboration = "TASSO",
    title = "{Evidence for Planar Events in $e^+ e^-$ Annihilation at High-Energies}",
    reportNumber = "DESY-79-53",
    doi = "10.1016/0370-2693(79)90830-X",
    journal = "Phys. Lett. B",
    volume = "86",
    pages = "243--249",
    year = "1979"
}

@article{ALEPH:2003obs,
    author = "Heister, A. and others",
    collaboration = "ALEPH",
    title = "{Studies of QCD at $e^+ e^-$ centre-of-mass energies between 91-GeV and 209-GeV}",
    reportNumber = "CERN-EP-2003-084",
    doi = "10.1140/epjc/s2004-01891-4",
    journal = "Eur. Phys. J. C",
    volume = "35",
    pages = "457--486",
    year = "2004"
}

@article{OPAL:2001klt,
    author = "Abbiendi, G. and others",
    collaboration = "OPAL",
    title = "{A Simultaneous measurement of the QCD color factors and the strong coupling}",
    eprint = "hep-ex/0101044",
    archivePrefix = "arXiv",
    reportNumber = "CERN-EP-2001-001",
    doi = "10.1007/s100520100699",
    journal = "Eur. Phys. J. C",
    volume = "20",
    pages = "601--615",
    year = "2001"
}

@article{Kluth:2000km,
    author = "Kluth, S. and Movilla Fernandez, P. A. and Bethke, S. and Pahl, C. and Pfeifenschneider, P.",
    title = "{A Measurement of the QCD color factors using event shape distributions at $\sqrt{s}$ = 14 GeV to 189 GeV}",
    eprint = "hep-ex/0012044",
    archivePrefix = "arXiv",
    reportNumber = "MPI-PHE-2000-19",
    doi = "10.1007/s100520100742",
    journal = "Eur. Phys. J. C",
    volume = "21",
    pages = "199--210",
    year = "2001"
}

@article{ALEPH:1997mcm,
    author = "Barate, R. and others",
    collaboration = "ALEPH",
    title = "{A Measurement of the QCD color factors and a limit on the light gluino}",
    reportNumber = "CERN-PPE-97-002, CERN-PPE-97-02, CERN-PPE-97-2, FSU-SCRI-97-126",
    doi = "10.1007/s002880050522",
    journal = "Z. Phys. C",
    volume = "76",
    pages = "1--14",
    year = "1997"
}

@article{ALEPH:1992fwh,
    author = "Decamp, D. and others",
    collaboration = "ALEPH",
    title = "{Evidence for the triple gluon vertex from measurements of the QCD color factors in Z decay into four jets}",
    reportNumber = "CERN-PPE-92-31",
    doi = "10.1016/0370-2693(92)91941-2",
    journal = "Phys. Lett. B",
    volume = "284",
    pages = "151--162",
    year = "1992"
}

@article{L3:1991jmb,
    author = "Adeva, B. and others",
    collaboration = "L3",
    title = "{A Test of QCD based on three jet events from Z0 decays}",
    reportNumber = "L3-030",
    doi = "10.1016/0370-2693(91)90504-J",
    journal = "Phys. Lett. B",
    volume = "263",
    pages = "551--562",
    year = "1991"
}

@article{L3:1990jlf,
    author = "Adeva, B. and others",
    collaboration = "L3",
    title = "{A Test of QCD based on four jet events from Z0 decays}",
    reportNumber = "L3-012",
    doi = "10.1016/0370-2693(90)90043-6",
    journal = "Phys. Lett. B",
    volume = "248",
    pages = "227--234",
    year = "1990"
}

@article{ALEPH:2002kjp,
    author = "Heister, A. and others",
    collaboration = "ALEPH",
    title = "{Measurements of the strong coupling constant and the QCD color factors using four jet observables from hadronic Z decays}",
    reportNumber = "CERN-EP-2002-029",
    doi = "10.1140/epjc/s2002-01114-2",
    journal = "Eur. Phys. J. C",
    volume = "27",
    pages = "1--17",
    year = "2003"
}

@article{Kluth:2006bw,
    author = "Kluth, Stefan",
    title = "{Tests of Quantum Chromo Dynamics at $e^+ e^-$ Colliders}",
    eprint = "hep-ex/0603011",
    archivePrefix = "arXiv",
    reportNumber = "MPP-2006-19",
    doi = "10.1088/0034-4885/69/6/R04",
    journal = "Rept. Prog. Phys.",
    volume = "69",
    pages = "1771--1846",
    year = "2006"
}

@article{Huston:2023ofk,
    author = "Huston, Joey and Rabbertz, Klaus and Zanderighi, Giulia",
    title = "{Quantum Chromodynamics}",
    eprint = "2312.14015",
    archivePrefix = "arXiv",
    primaryClass = "hep-ph",
    month = "12",
    year = "2023",
    journal = ""
}

@article{Sherpa:2019gpd,
    author = "Bothmann, Enrico and others",
    collaboration = "Sherpa",
    title = "{Event Generation with Sherpa 2.2}",
    eprint = "1905.09127",
    archivePrefix = "arXiv",
    primaryClass = "hep-ph",
    reportNumber = "FERMILAB-PUB-19-218-T, SLAC-PUB-17433, IPPP/19/42, MCNET-19-11",
    doi = "10.21468/SciPostPhys.7.3.034",
    journal = "SciPost Phys.",
    volume = "7",
    number = "3",
    pages = "034",
    year = "2019"
}

@article{deBlas:2025gyz,
    author = "de Blas, Jorge and others",
    title = "{Physics Briefing Book: Input for the 2026 update of the European Strategy for Particle Physics}",
    eprint = "2511.03883",
    archivePrefix = "arXiv",
    primaryClass = "hep-ex",
    reportNumber = "CERN--2025-008, CERN-ESU-2025-001",
    doi = "10.17181/CERN.35CH.2O2P",
    month = "11",
    year = "2025",
    journal = {}
}

@article{FCC:2018byv,
    author = "Abada, A. and others",
    collaboration = "FCC",
    title = "{FCC Physics Opportunities}: {Future Circular Collider Conceptual Design Report Volume 1}",
    reportNumber = "CERN-ACC-2018-0056",
    doi = "10.1140/epjc/s10052-019-6904-3",
    journal = "Eur. Phys. J. C",
    volume = "79",
    number = "6",
    pages = "474",
    year = "2019"
}

@article{FCC:2018evy,
    author = "Abada, A. and others",
    collaboration = "FCC",
    title = "{FCC-ee: The Lepton Collider}: {Future Circular Collider Conceptual Design Report Volume 2}",
    reportNumber = "CERN-ACC-2018-0057",
    doi = "10.1140/epjst/e2019-900045-4",
    journal = "Eur. Phys. J. ST",
    volume = "228",
    number = "2",
    pages = "261--623",
    year = "2019"
}

@article{FCC:2025lpp,
    author = "Benedikt, M. and others",
    collaboration = "FCC",
    title = "{Future Circular Collider Feasibility Study Report: Volume 1, Physics, Experiments, Detectors}",
    eprint = "2505.00272",
    archivePrefix = "arXiv",
    primaryClass = "hep-ex",
    reportNumber = "CERN-FCC-PHYS-2025-0002",
    doi = "10.1140/epjc/s10052-025-15077-x",
    journal = "Eur. Phys. J. C",
    volume = "85",
    number = "12",
    pages = "1468",
    year = "2025"
}

@article{An:2018dwb,
    author = "An, Fenfen and others",
    title = "{Precision Higgs physics at the CEPC}",
    eprint = "1810.09037",
    archivePrefix = "arXiv",
    primaryClass = "hep-ex",
    reportNumber = "FERMILAB-PUB-18-573-T",
    doi = "10.1088/1674-1137/43/4/043002",
    journal = "Chin. Phys. C",
    volume = "43",
    number = "4",
    pages = "043002",
    year = "2019"
}

@inproceedings{CEPCPhysicsStudyGroup:2022uwl,
    author = "Cheng, Huajie and others",
    collaboration = "CEPC Physics Study Group",
    title = "{The Physics potential of the CEPC. Prepared for the US Snowmass Community Planning Exercise (Snowmass 2021)}",
    booktitle = "{Snowmass 2021}",
    eprint = "2205.08553",
    archivePrefix = "arXiv",
    primaryClass = "hep-ph",
    month = "5",
    year = "2022",
    journal = {}
}

@article{Ai:2024nmn,
    author = "Ai, Xiaocong and others",
    title = "{Flavor Physics at the CEPC: a General Perspective}",
    eprint = "2412.19743",
    archivePrefix = "arXiv",
    primaryClass = "hep-ex",
    doi = "10.1088/1674-1137/adf1f0",
    journal = "Chin. Phys.",
    volume = "49",
    number = "10",
    pages = "103003",
    year = "2025"
}

@article{Electron-PositronAlliance:2019cpi,
    author = "Badea, Anthony and Baty, Austin and Chang, Paoti and Innocenti, Gian Michele and Maggi, Marcello and Mcginn, Christopher and Peters, Michael and Sheng, Tzu-An and Thaler, Jesse and Lee, Yen-Jie",
    collaboration = "Electron-Positron Alliance",
    title = "{Measurements of two-particle correlations in $e^+e^-$ collisions at 91 GeV with ALEPH archived data}",
    eprint = "1906.00489",
    archivePrefix = "arXiv",
    primaryClass = "hep-ex",
    reportNumber = "MITHIG-MOD-19-001",
    doi = "10.1103/PhysRevLett.123.212002",
    journal = "Phys. Rev. Lett.",
    volume = "123",
    number = "21",
    pages = "212002",
    year = "2019"
}

@article{Nagy:1997yn,
    author = "Nagy, Zoltan and Trocsanyi, Zoltan",
    title = "{Next-to-leading order calculation of four jet shape variables}",
    eprint = "hep-ph/9707309",
    archivePrefix = "arXiv",
    doi = "10.1103/PhysRevLett.79.3604",
    journal = "Phys. Rev. Lett.",
    volume = "79",
    pages = "3604--3607",
    year = "1997"
}

@article{Catani:1998bh,
    author = "Catani, Stefano",
    title = "{The Singular behavior of QCD amplitudes at two loop order}",
    eprint = "hep-ph/9802439",
    archivePrefix = "arXiv",
    reportNumber = "CERN-TH-98-42, LPTHE-ORSAY-97-57",
    doi = "10.1016/S0370-2693(98)00332-3",
    journal = "Phys. Lett. B",
    volume = "427",
    pages = "161--171",
    year = "1998"
}

@article{Sterman:2002qn,
    author = "Sterman, George F. and Tejeda-Yeomans, Maria E.",
    title = "{Multiloop amplitudes and resummation}",
    eprint = "hep-ph/0210130",
    archivePrefix = "arXiv",
    reportNumber = "YITP-SB-02-56",
    doi = "10.1016/S0370-2693(02)03100-3",
    journal = "Phys. Lett. B",
    volume = "552",
    pages = "48--56",
    year = "2003"
}

@article{Becher:2009cu,
    author = "Becher, Thomas and Neubert, Matthias",
    title = "{Infrared singularities of scattering amplitudes in perturbative QCD}",
    eprint = "0901.0722",
    archivePrefix = "arXiv",
    primaryClass = "hep-ph",
    reportNumber = "FERMILAB-PUB-09-002-T, MZ-TH-09-01",
    doi = "10.1103/PhysRevLett.102.162001",
    journal = "Phys. Rev. Lett.",
    volume = "102",
    pages = "162001",
    year = "2009",
    note = "[Erratum: Phys.Rev.Lett. 111, 199905 (2013)]"
}

@article{Abreu:2021oya,
    author = "Abreu, S. and Febres Cordero, F. and Ita, H. and Page, B. and Sotnikov, V.",
    title = "{Leading-color two-loop QCD corrections for three-jet production at hadron colliders}",
    eprint = "2102.13609",
    archivePrefix = "arXiv",
    primaryClass = "hep-ph",
    reportNumber = "CERN-TH-2021-023, FR-PHENO-2021-06, MPP-2021-20",
    doi = "10.1007/JHEP07(2021)095",
    journal = "JHEP",
    volume = "07",
    pages = "095",
    year = "2021"
}

@article{Campbell:2016tcu,
    author = "Campbell, John M. and Ellis, R. Keith",
    title = "{Top-quark loop corrections in Z+jet and Z + 2 jet production}",
    eprint = "1610.02189",
    archivePrefix = "arXiv",
    primaryClass = "hep-ph",
    reportNumber = "FERMILAB-PUB-16-429-T, IPPP-16-88",
    doi = "10.1007/JHEP01(2017)020",
    journal = "JHEP",
    volume = "01",
    pages = "020",
    year = "2017"
}

@article{Nagy:1998kw,
    author = "Nagy, Zoltan and Trocsanyi, Zoltan",
    editor = "Narison, Stephan",
    title = "{Multijet rates in e+ e- annihilation: Perturbation theory versus LEP data}",
    eprint = "hep-ph/9808364",
    archivePrefix = "arXiv",
    reportNumber = "CERN-TH-98-266",
    doi = "10.1016/S0920-5632(99)00130-9",
    journal = "Nucl. Phys. B Proc. Suppl.",
    volume = "74",
    pages = "44--48",
    year = "1999"
}

@article{Baberuxki:2019ifp,
    author = "Baberuxki, Nick and Preuss, Christian T. and Reichelt, Daniel and Schumann, Steffen",
    title = "{Resummed predictions for jet-resolution scales in multijet production in e$^{+}$e$^{-}$ annihilation}",
    eprint = "1912.09396",
    archivePrefix = "arXiv",
    primaryClass = "hep-ph",
    reportNumber = "MCNET-19-29, FERMILAB-PUB-19-639-T",
    doi = "10.1007/JHEP04(2020)112",
    journal = "JHEP",
    volume = "04",
    pages = "112",
    year = "2020"
}
\end{document}